\documentclass[10pt,journal,twocolumn]{IEEEtran}
\usepackage{multirow}
\usepackage{amssymb}
\usepackage[dvips]{graphicx}
\usepackage{amsthm}
\usepackage{latexsym,bm}
\usepackage{color}
\usepackage{subfigure}
\usepackage{lettrine}
\usepackage{multirow}
\usepackage{amsmath}
\usepackage{longtable}
\usepackage{accents}
\usepackage{cite}
\usepackage{enumerate}
\usepackage{arydshln}
\theoremstyle{plain}

\theoremstyle{definition}
\theoremstyle{definition}

\begin{document}

\title{QoS-Aware Buffer-Aided Relaying Implant WBAN for Healthcare IoT: Opportunities and Challenges
\thanks{Guofa Cai, Yi Fang (corresponding author), and Guojun Han are with the Guangdong University of Technology and also with the State Key Laboratory of Integrated Services Networks, Xidian University; Jinming Wen is with the Jinan University; Xiaodong Yang is with the Xidian University.}
}

\author{Guofa Cai, Yi Fang, Jinming Wen, Guojun Han, and Xiaodong Yang}


\maketitle

\begin{abstract}
Internet of Things (IoT) have motivated a paradigm shift in the development of various applications such as mobile health.
Wireless body area network (WBAN) comprises many low-power devices in, on, or around the human body, which offers a desirable solution to monitor physiological signals for mobile-health applications.
In the implant WBAN, an implant medical device transmits its measured biological parameters to a target hub with the help of at least one on-body device(s) to satisfy its strict requirements on size, quality of service (QoS, e.g., reliability), and power consumption.
In this article, we first review the recent advances of conventional cooperative WBAN.
Afterwards, to address the drawbacks of the conventional cooperative WBAN, a QoS-aware buffer-aided relaying framework is proposed for the implant WBAN.
In the proposed framework, hierarchical modulations are considered to fulfill the different QoS requirements of different sensor data from an implant medical device.
We further conceive some new transmission strategies for the buffer-aided signal-relay and multi-relay implant WBANs. Simulation results show that the proposed cooperative WBAN provides better performance than the conventional cooperative counterparts. Finally, some open research challenges regarding the buffer-aided multi-relay implant WBAN are pointed out to inspire more research activities.



\end{abstract}



\section{Introduction}
\IEEEPARstart{P}{OPULATION} aging has been sweeping around the world. The problems of chronic diseases, medical quality and the strained healthcare infrastructure have become an important challenge for the long-term development of the world.
To improve the medical quality and reduce the healthcare cost, it is a matter of the utmost urgency to use information technologies to solve the current medical problems.
Healthcare Internet of Things (IoT) have facilitated the development of mobile-health systems that support gathering, delivery, and retrieval of medical information \cite{1-1,1-2}.
Recently, wireless body area network (WBAN), which has been reported in IEEE 802.15.6 \cite{1}, is considered as a new technology for mobile-health applications to implement the medical devices either implanted in or placed on/around the human body.

\begin{figure*}
\centering
\subfigure[]{ \label{fig:subfig:1a} 
\includegraphics[width=2.2in,height=1.6in]{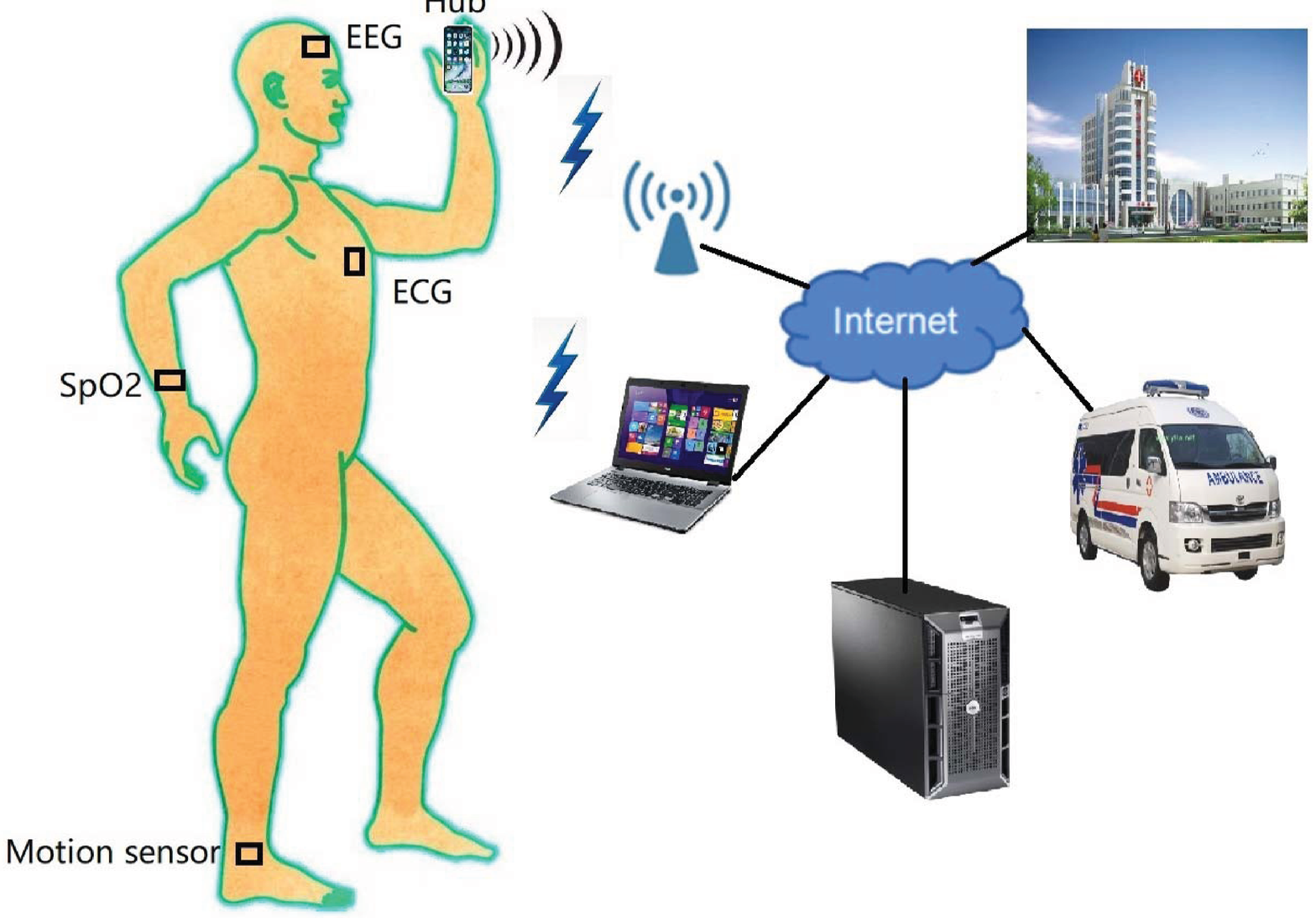}}
\hspace{1in}
\subfigure[]{ \label{fig:subfig:1b} 
\includegraphics[width=2.2in,height=1.6in]{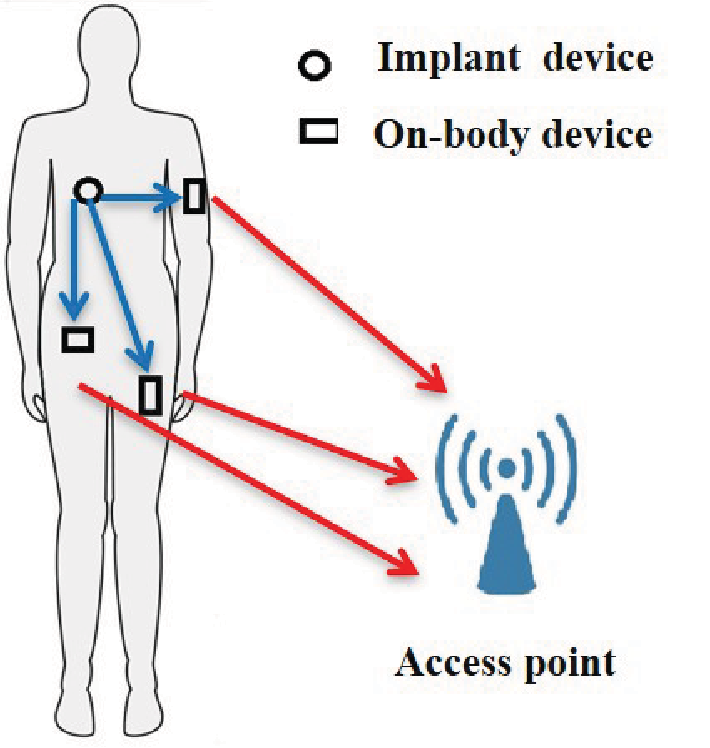}}
\vspace{-1.5mm}
\caption{Frameworks of (a) a typical WBAN and (b) a conventional cooperative WBAN.}
\label{fig:Fig.1} 
\end{figure*}

A typical WBAN framework, as shown in Fig.~\ref{fig:subfig:1a}, consists of many low-power medical devices to measure biological parameters from the human body, e.g., emergency electroencephalogram
(EEG) and electrocardiogram (ECG) data. More specifically, these useful biological parameters are first collected in an on-body device (i.e., the target hub), e.g., personal digital assistant (PDA), and the patient can know his/her condition in real time. After that, the target hub transmits these parameters to the remote off-body devices (e.g., monitoring central) through mobile communication network or Internet. In this sense, the doctor can observe the patient's condition in time and provide the patient with precise treatment.
Unlike through-the-air wireless communications, the implant medical devices encounter many challenges because living tissues are considered as part of its transmission channel \cite{2}.
First, since the human body is a hostile channel to high frequency electromagnetic signals, the signals propagation is attenuated considerably.
Second, the implant devices should guarantee stringent-miniaturisation and quality-of-service (QoS) provisioning due to the invasive nature of implantation surgeries.

Among all the challenges, how to guarantee the QoS for medical applications is considered as the most important issue \cite{3}.
It is obvious that equal-importance data transmission for the implant WBAN doesn't satisfy different classes of medical data, which includes one class of emergent data and multiple classes of non-emergent data.
The emergent medical data should be transmitted in a highest-priority (HP) level, while the non-emergent medical data should be differentiated according to different application purposes.
For instance, the implant device, \textit{Pill Cam}, can get the gastrointestinal video by using a camera, where the video includes a base layer and enhancement layer data streams. The base layer corresponding to the HP data and  the enhancement layer corresponding to the low-priority (LP) data are recognized as the emergent and non-emergent data, respectively. \textit{Pill Cam} also needs to obtain other non-emergent data, e.g., temperature. It is intuitive that the LP data is more critical than the temperature.
Hence, it is necessary to transmit the medical data with different priorities to satisfy the diverse performance requirements.
In addition to \textit{Pill Cam}, the medical information from other implant devices, such as intra-cranial pressure monitors, glucose sensors, deep brain activity sensors, oximeters and pH sensors, should be also transmitted by using different priorities.


To realize high-reliability medical data transmission with different QoS, cooperative technologies have been introduced into the implant WBAN \cite{4, 5}.
A conventional cooperative implant WBAN framework, as shown in Fig.~\ref{fig:subfig:1b}, has been extensively studied. To be specific, an implant device (source) communicates with an off-body access point (destination) via a number of on-body devices (relays), where decode-and-forward (DF) and amplify-and-forward (AF) relaying strategies are commonly used.
For DF strategy, the relay decodes the source information from its received signal, re-encodes and forwards it to the destination; while for AF strategy, the relay amplifies its received signal and immediately transmits it to the destination.
However, the performance of the cooperative WBAN with both DF and AF strategies is limited by the worst component channel among all links.

To tackle the aforementioned weakness, the transmission nodes in the implant WBAN system can be selected opportunistically based on the channel link quality \cite{6}\footnote{In the reference \cite{6}, the link-quality estimate methods in WBAN can be categorized as hardware-based, software-based and inertial sensor-based methods. For convenience, we assume that the channel state information (CSI) can be acquired using the above methods.}, and thus better performance is expected compared with the conventional cooperative WBAN.
This opportunistic scheduling requires buffers at the relays such that the received information can be stored until good channel quality is observed \cite{7}.
In the implant WBAN, the relay can incorporate a relative-large-size and  miniaturized buffer with the development of modern data-storage technology.
Moreover, although buffer-aided relaying significantly improves the performance of WBAN \cite{8}, it suffers from some practical drawbacks, e.g., introducing an additional delay and an increased complexity of the relays.
As is well known, hierarchical modulation is a promising technique to provide different QoS \cite{9-1}. Using hierarchical modulation, a bit stream can be separated into several multiplexed sub-streams with different priority levels, i.e., HP and LP sub-streams. Hence, it is intuitive that the hierarchical modulation can be exploited to satisfy different QoS requirements for implant WBAN from the physical-layer perspective.
In addition, although buffer-aided relaying is not suitable for delay-sensitive information transmission in the implant WBAN, the employment of hierarchical modulation can mitigate the additional delay.
Especially, the delay-sensitive information can be mapped to the HP layer of the hierarchical-modulation symbols so as to achieve lower delay, while the delay-tolerant information can be mapped to the LP layer. Therefore, the proposed buffer-aided relaying WBAN can transmit simultaneously the delay-sensitive and delay-tolerant information to mitigate the delay and achieve desirable performance gain.

The goal of this article is to present a QoS-aware buffer-aided relaying framework for the implant WBAN.
Apart from the system framework, some new transmission strategies for the buffer-aided single- and multi-relay WBANs are proposed and their corresponding performance is investigated in terms of bit error rate (BER) and system delay.
Simulation results illustrate that the proposed cooperative WBAN provides better BER performance than the conventional counterparts.
Last but not least, we discuss several interesting open problems, which are of great importance to further reduce the delay and improve error performance of the proposed implant WBAN.

The rest of the article is organized as follows. In Sect.~II, we review the recent advances of conventional cooperative WBAN and present a buffer-aided relaying implant WBAN framework. In Sects.~II and III, we propose some new transmission strategies for buffer-aided single- and multi-relay WBANs and investigate their performance. In Sects.~IV and V, we provide some future research topics and conclude this work, respectively.

\section{Cooperative Communication for Implant WBAN}
In this section, we first review the recent advances of conventional cooperative WBAN system. Then we propose a QoS-aware buffer-aided relaying framework for the implant WBAN.

\subsection{Conventional Cooperative WBAN}
As a new technology for healthcare IoT, a WBAN usually consists of a number of battery-driven medical devices, which are either placed on/around or implanted in the human body. The IEEE 802.15.6 standard for WBAN supports two physical-layers: Narrowband (NB) and Ultra-wideband (UWB). In a WBAN, both NB and UWB techniques can be utilized to implement these medical devices. For the above two techniques, designing low-power and high-reliability transmission schemes is a very important issue due to the capacity-limited batteries and life-vital information.

Cooperative communication can improve link reliability and save power consumption due to the advantage of spatial diversity.
In a cooperative UWB-based WBAN framework, the link-reliability analysis has been first introduced in \cite{9}, where a two-relay-assisted transmission scenario is considered.
Under the same framework, a joint relay-selection and power-control strategy with QoS provisioning has been proposed by using a game-theoretic approach, which reduces power consumption at the expense of the delay \cite{10}.
Noticeably, the above works are all related to on-body device-aided cooperative WBAN. As mentioned previously, implant devices need to provide reliability, low-power consumption, and long-lifetime transmission under great path-loss and deep fading environments.

With these motivations, a cooperative implant WBAN framework has been proposed in \cite{4, 5, 13}, where multiple on-body devices are employed to assist an implant device to communicate with an access point to realize reliable transmission.
More specifically, an incremental relay-based cooperative routing protocol has been proposed for the implant WBAN to reduce the energy consumption of the implant device and satisfy a flexible QoS requirement \cite{13}. Differentiating from the perspective of the media access control (MAC) layer in \cite{13}, the closed-form expressions of the average BER and outage probability of the cooperative implant WBAN have been evaluated over realistic human-body channels, especially considering the specific binary PSK modulation \cite{4}.
In addition, the authors have also discussed the tradeoff between the required transmission power and QoS requirement (i.e., average BER or outage probability) so as to facilitate the design of cooperative implant WBAN.
Furthermore, in cooperative UWB-based implant WBAN, an distributed beamforming strategy has been developed to achieve low power consumption \cite{5}. In such a network, the relay location has a significant impact on the system performance.

Through the above discussions, we can make the following three observations: 1) Most existing literature on WBAN only considers how to fulfil different QoS requirements for different types of information from the perspective of the MAC layer; 2) Although the QoS requirement has been investigated in the cooperative WBAN by using some specific modulations, all transmitted bits are viewed as the equal-importance bits; 3) The performance of the conventional cooperative WBAN is limited by the worst component channel among all channel links. Thereby, in the next subsection, we propose a novel QoS-aware buffer-aided relaying framework for the implant WBAN to significantly improve the performance gain.

\subsection{Buffer-Aided Relaying Implant WBAN Framework}
\begin{figure}[t] 
\center\vspace{-2mm}
\includegraphics[width=2.6in,height=1.8in]{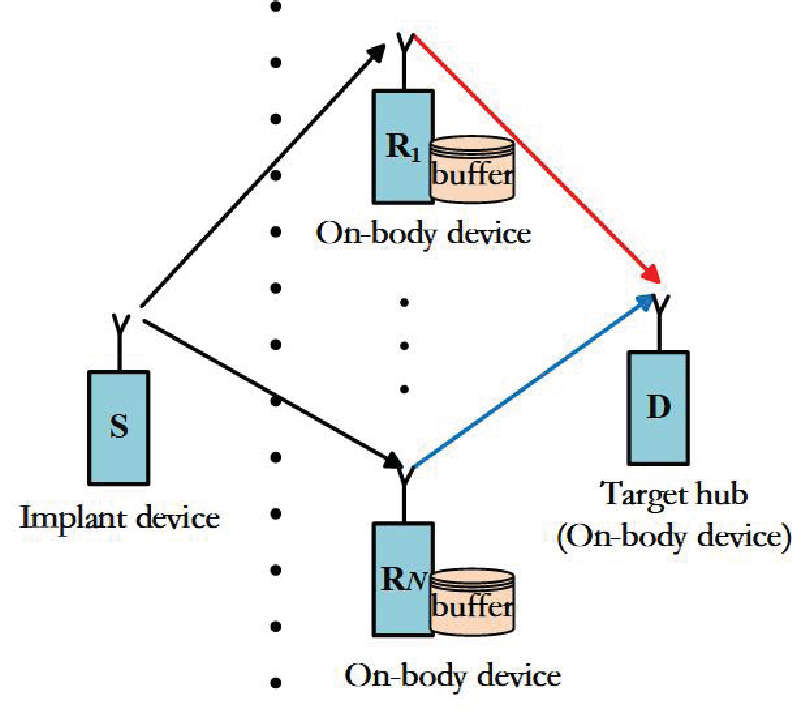}
\vspace{-2mm}
\caption{Buffer-aided relaying framework for the implant WBAN.}
\label{fig:Fig.2}\vspace{-2mm}
\end{figure}

As illustrated in Fig.~\ref{fig:Fig.2}, the proposed buffer-aided relaying framework for the implant WBAN consists a source node $\textbf{S}$ (i.e., the implant device), $N$ relay nodes $\textbf{R}_n$ (i.e., the on-body devices), and a destination node $\textbf{D}$ (i.e., the on-body device or the target hub), where a direct source-to-destination link is not available in practical WBAN applications and the DF protocol is employed at the relay nodes to forward the signals.
First, hierarchical PSK (HPSK) modulation is adopted by all nodes because it has a constant envelope and low peak to average power ratio \cite{9-1}. In the WBAN, using HPSK modulation can satisfy different QoS requirements (i.e., BER) for different bits, where the phase parameter $\bm{\theta}$ can be adjusted \cite{9-1}. In a HPSK-modulation symbol, the delay-sensitive data are mapped to the HP bits while the delay-tolerant ones are mapped to the LP bits, i.e., both types of data service can be transmitted simultaneously, thus mitigating the entire delay of the proposed framework.
Second, it is assumed that each node is equipped with single antenna and operates in half-duplex mode.
Moreover, each relay node has an additional data buffer of size $L$.
The number of packets stored in the $n$-th relay buffer is increased by one when a packet is successfully received by the $n$-th relay node while it is decreased by one when a packet is successfully transmitted by the $n$-th relay node, which is smaller than the buffer size.
Third, the in-body-to-on-body and on-body-to-on-body channel models in \cite{15} are considered as the channels for the source-to-relay ($\textbf{SR}_n$) and the relay-to-destination ($\textbf{R}_n \textbf{D}$), respectively.
Furthermore, we assume that the channel coefficients are constant during each time slot and change independently from one time slot to another.
We also assume that the power for the additive white Gaussian noise is $-117.73$~dBm and the devices are working at Industrial, Scientific and Medical band.
It should be noted that the proposed system framework practically works in time-division duplexing (TDD) mode, thus the channel reciprocity can be well exploited.

Based on the link quality (i.e., CSI) and buffer status information of the relays, the relay node is selected to either transmit or receive a packet. The basic principle of the transmission protocol in the proposed framework is described as follows: $1)$~If the buffer is full (i.e., only the $\textbf{R}_n \textbf{D}$ link is available), the $n$-th relay node can transmit a packet to the destination node over the $\textbf{R}_n \textbf{D}$ link;
$2)$~If the buffer is empty (i.e., only the $\textbf{S} \textbf{R}_n$ link is available), the $n$-th relay node can receive a packet from  the source node over the $\textbf{S} \textbf{R}_n$ link;
$3)$~If the buffer is neither empty nor full (i.e., both the $\textbf{R}_n \textbf{D}$  and $\textbf{S} \textbf{R}_n$ links are available), the $n$-th relay node can either transmit a packet to the destination node over the $\textbf{R}_n \textbf{D}$ link or receive a packet from the source node over the $\textbf{S} \textbf{R}_n$ link.

Compared with the on-body-to-on-body channel, the in-body-to-on-body channel has a higher path loss and shadowing due to different material dielectric properties inside the human body. Many previous transmission protocols and results for the conventional wireless communications are not suitable for the proposed framework \cite{7}. For this reason, some novel protocols for the proposed framework are worth to be investigated.

\begin{figure}[t] 
\center\vspace{-2mm}
\includegraphics[width=3.5in,height=3.1in]{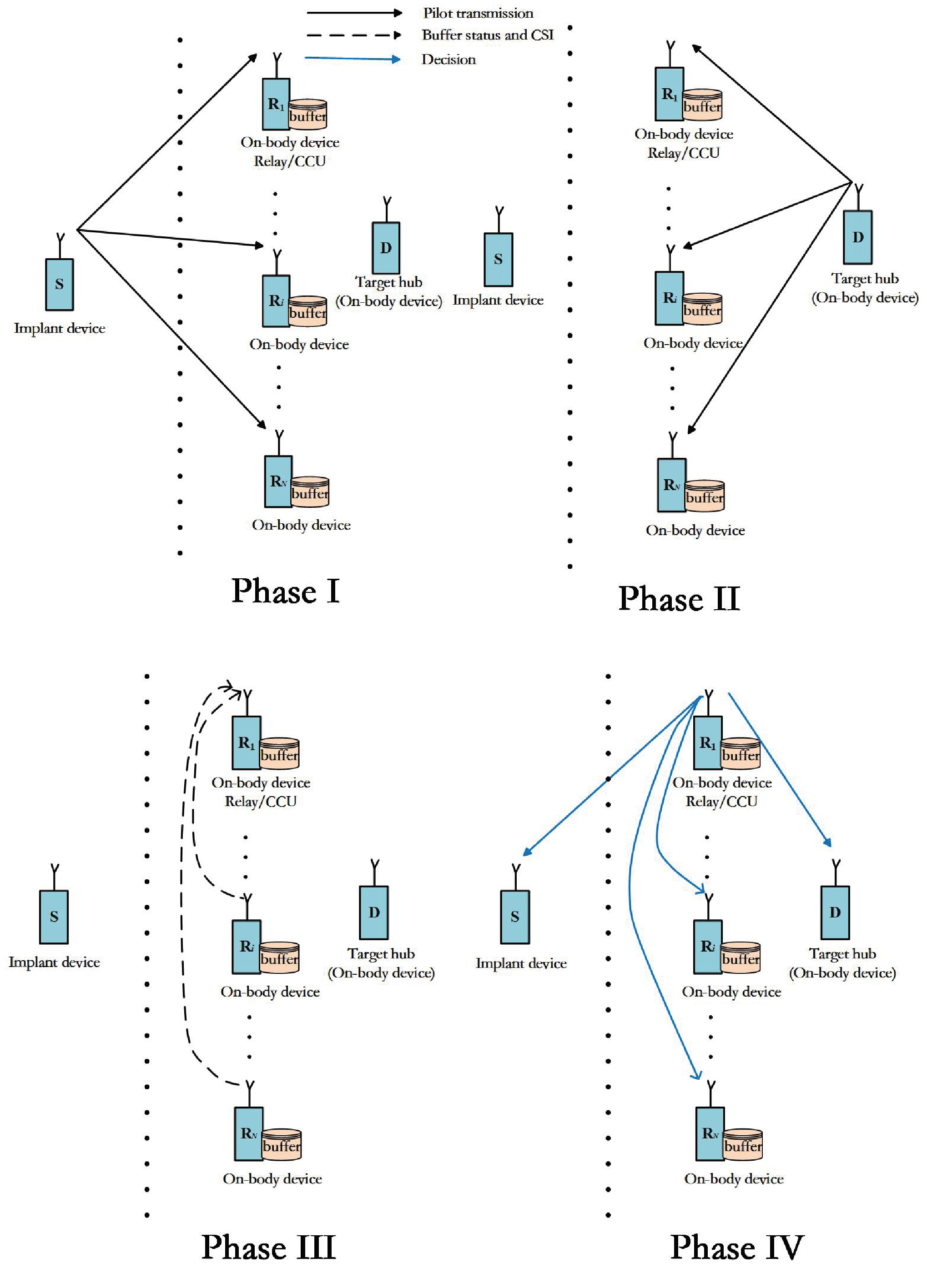}
\vspace{-2mm}
\caption{The four phases of CSI and buffer status exchange.}
\label{fig:Fig.2-1}\vspace{-2mm}
\end{figure}

\subsection{Implementation of Relay/Link Selection}
In practical applications, both centralized and distributed coordinations \cite{7} can be used to select the best relay in the proposed framework. In the distributed coordination, a central control unit (CCU) is used to acquire the link-quality information of all links and the buffer sizes for all relays. In the proposed framework, it is obvious that a relay node is more likely to serve as the CCU with respect to the source and destination nodes. Generally speaking, the relay with CCU may require stronger processing capability than the other relays, thus this node should be carefully designed.
In this paper, the detailed implementation for the proposed framework is described as follows.

For the multi-relay case, to obtain the best relay node, we divide the relay-selection process into the four phases,  which is shown in Fig.~\ref{fig:Fig.2-1}.\footnote{
Although the timing-synchronization design and its performance analysis for the proposed framework are very interesting, the investigation is beyond the scope of this tutorial article, which deserves another article in the future.}
In phase I, the source node broadcasts pilot symbols and each relay estimates its CSI of each $\textbf{S} \textbf{R}$ link.
In phase II, the destination node broadcasts pilot symbols and each relay estimates its CSI of each $\textbf{R}_n \textbf{D}$ link.
In phase III, except for the CCU/relay, the other relay sends the buffer status information and CSI to the CCU/relay.
In phase IV, the CCU/relay decides which relay can transmit, receive or stay idle according to the CSIs and buffer status information from all links.
However, for the single-relay case, the relay-selection process only includes three phases because the phase III is not necessary to be performed.

In the above implementation, a general method is given to obtain global CSI and buffer status information. However, the number of channel estimations depends on the relay/link-selection strategy. Hence, with an aim to reduce the overhead energy consumption, we should design a good relay/link-selection strategy.

\begin{figure*}
\centering
\subfigure[Three transmission scenarios]{ \label{fig:subfig:3a} 
\includegraphics[width=3.1in,height=1.7in]{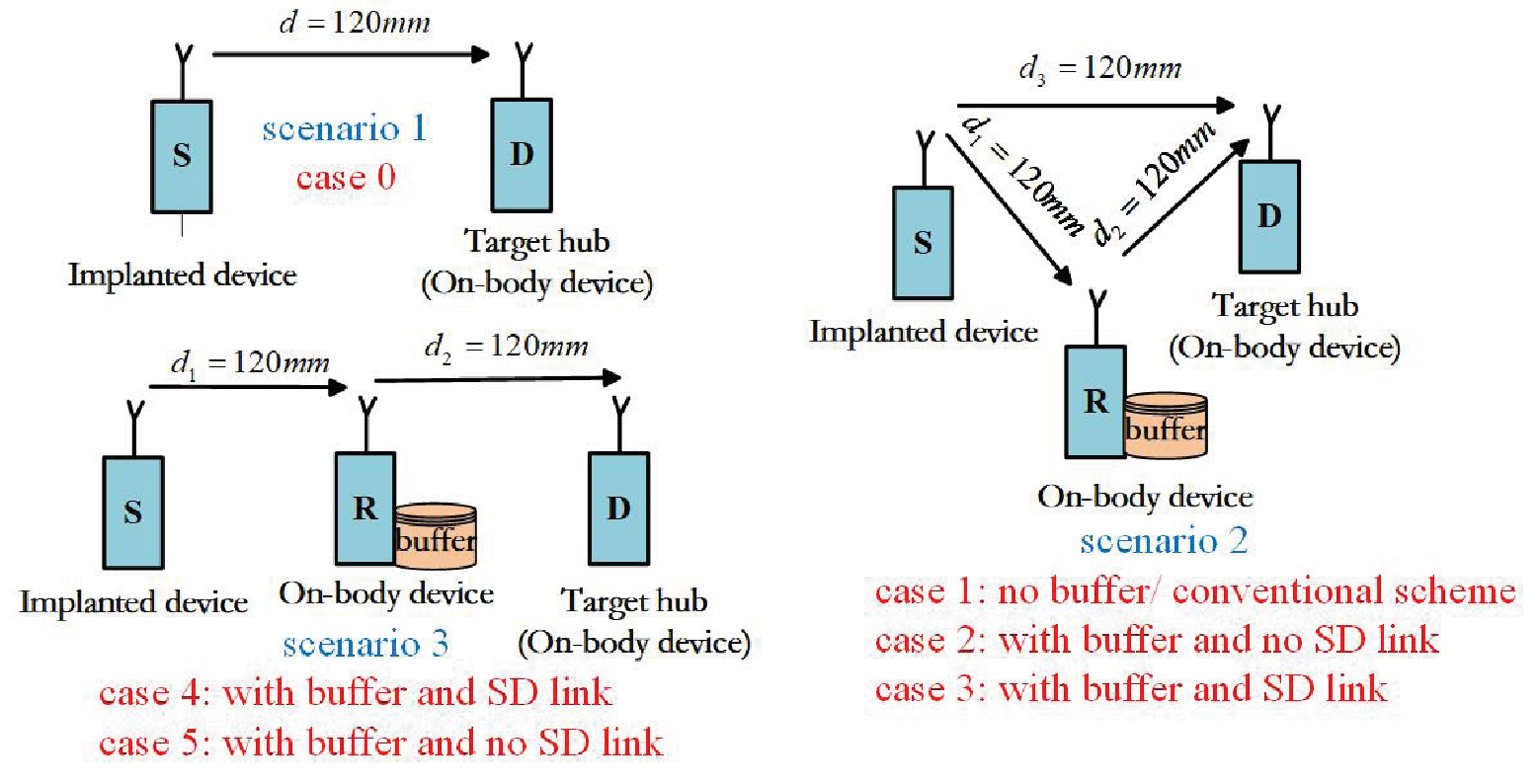}}
\subfigure[BER performance of different cases]{ \label{fig:subfig:3b} 
\includegraphics[width=1.9in,height=1.7in]{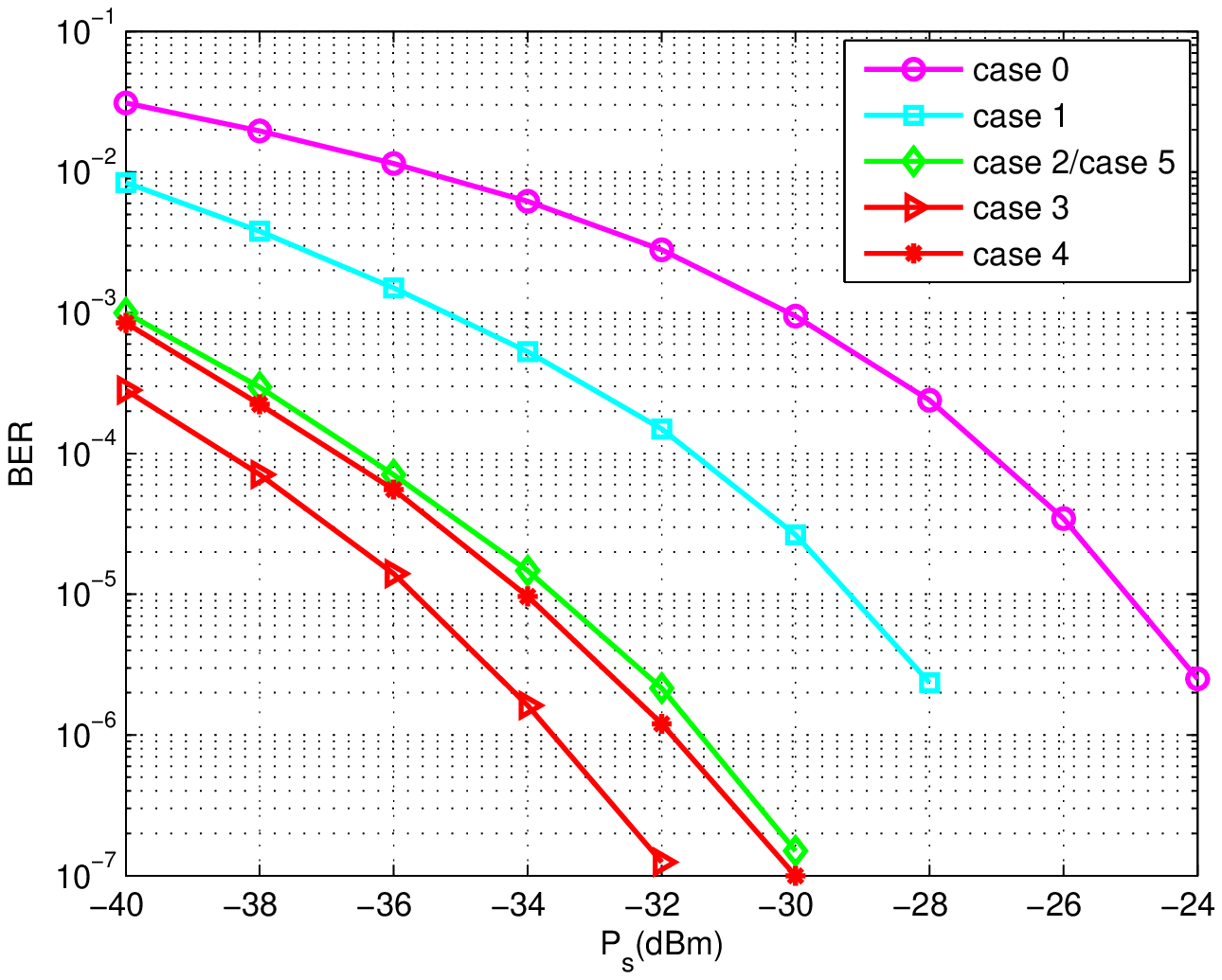}}
\subfigure[The effect of $\beta$ and $L$ on the BER performance of Protocol~1]{ \label{fig:subfig:3d} 
\includegraphics[width=1.9in,height=1.7in]{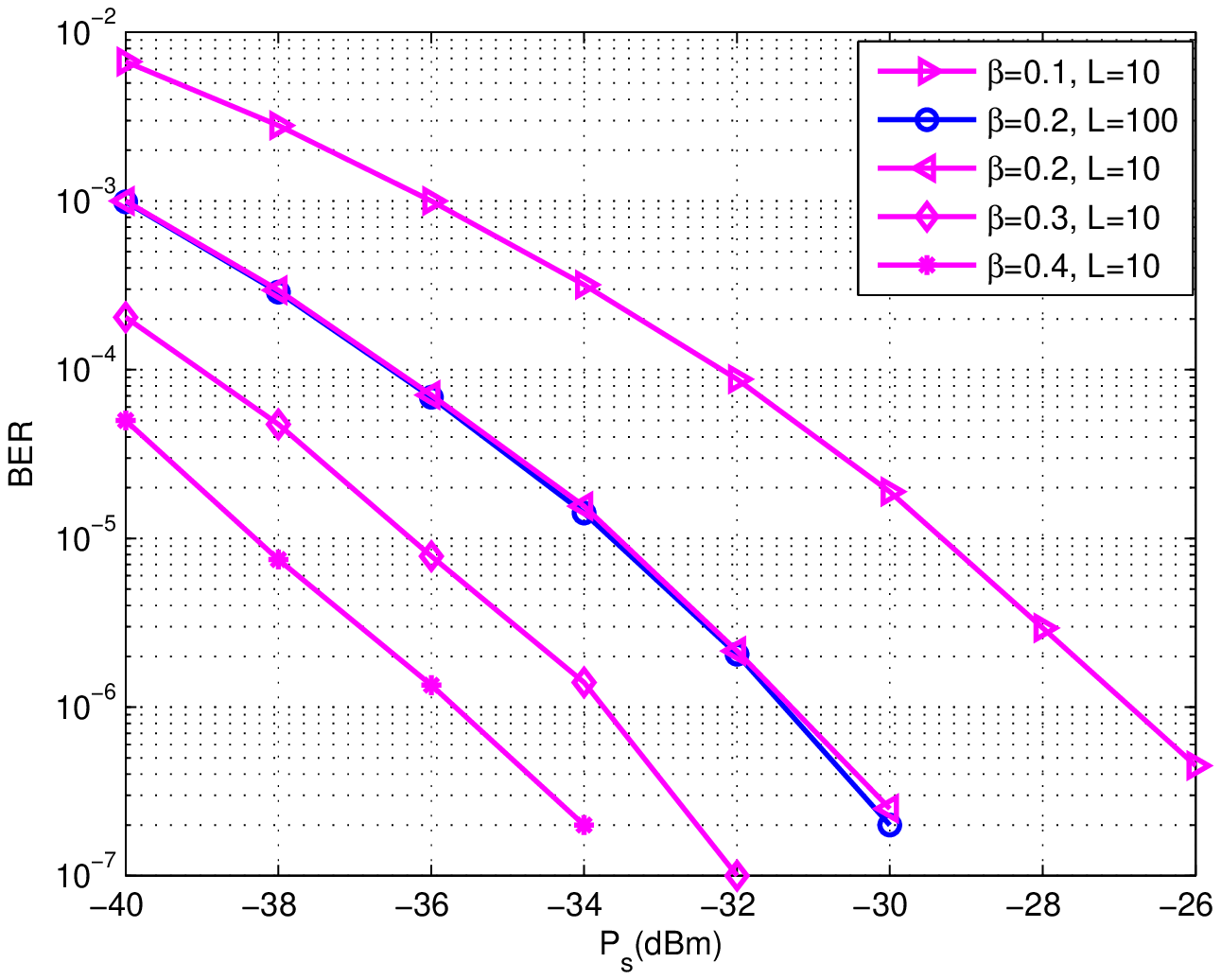}}
\subfigure[The effect of $\beta$ and $L$ on the BER performance of Protocol~$1^{*}$]{ \label{fig:subfig:3d-1} 
\includegraphics[width=2.2in,height=1.8in]{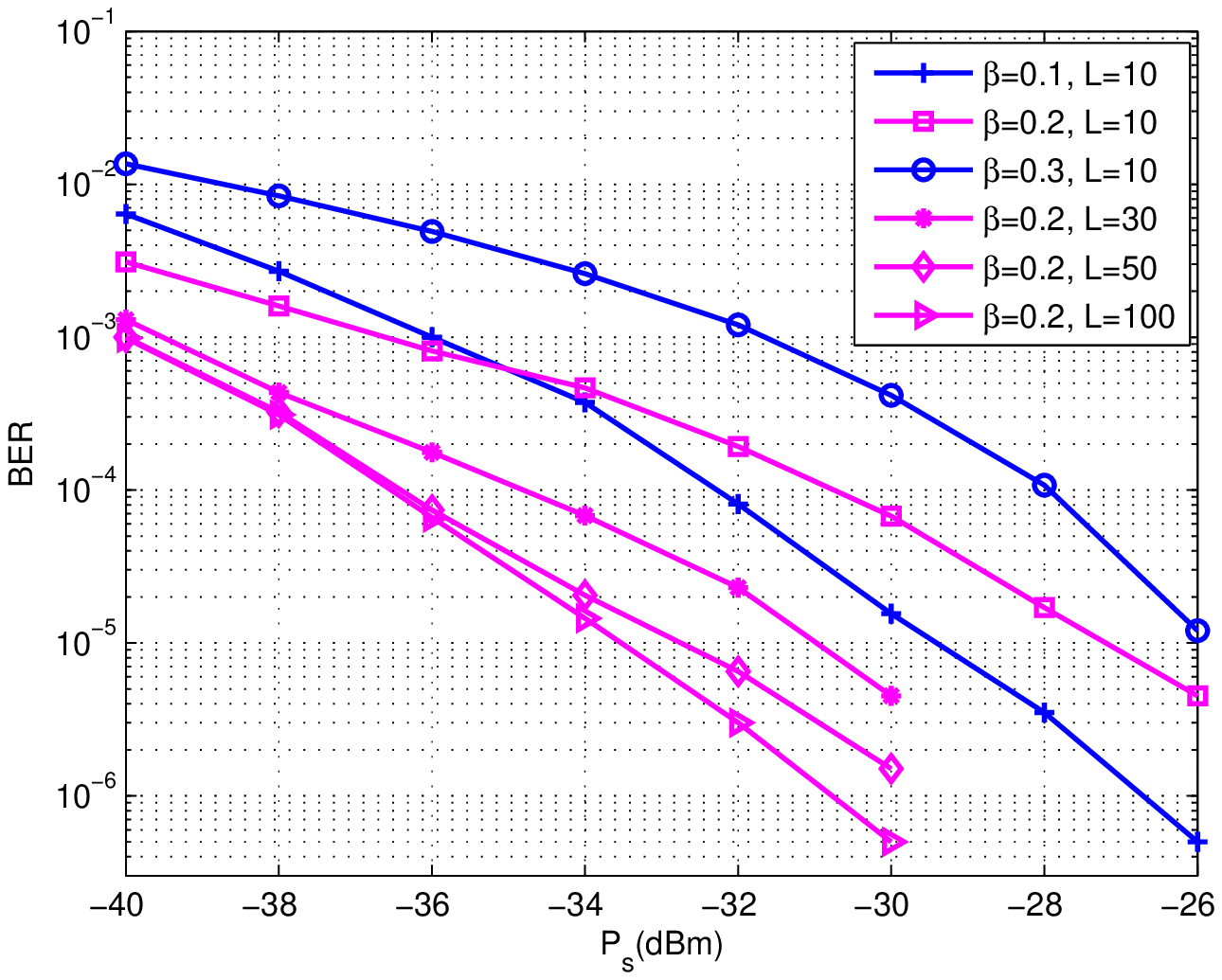}}
\subfigure[Average system delay versus $\beta$ of different protocols]{ \label{fig:subfig:3d-2} 
\includegraphics[width=2.2in,height=1.8in]{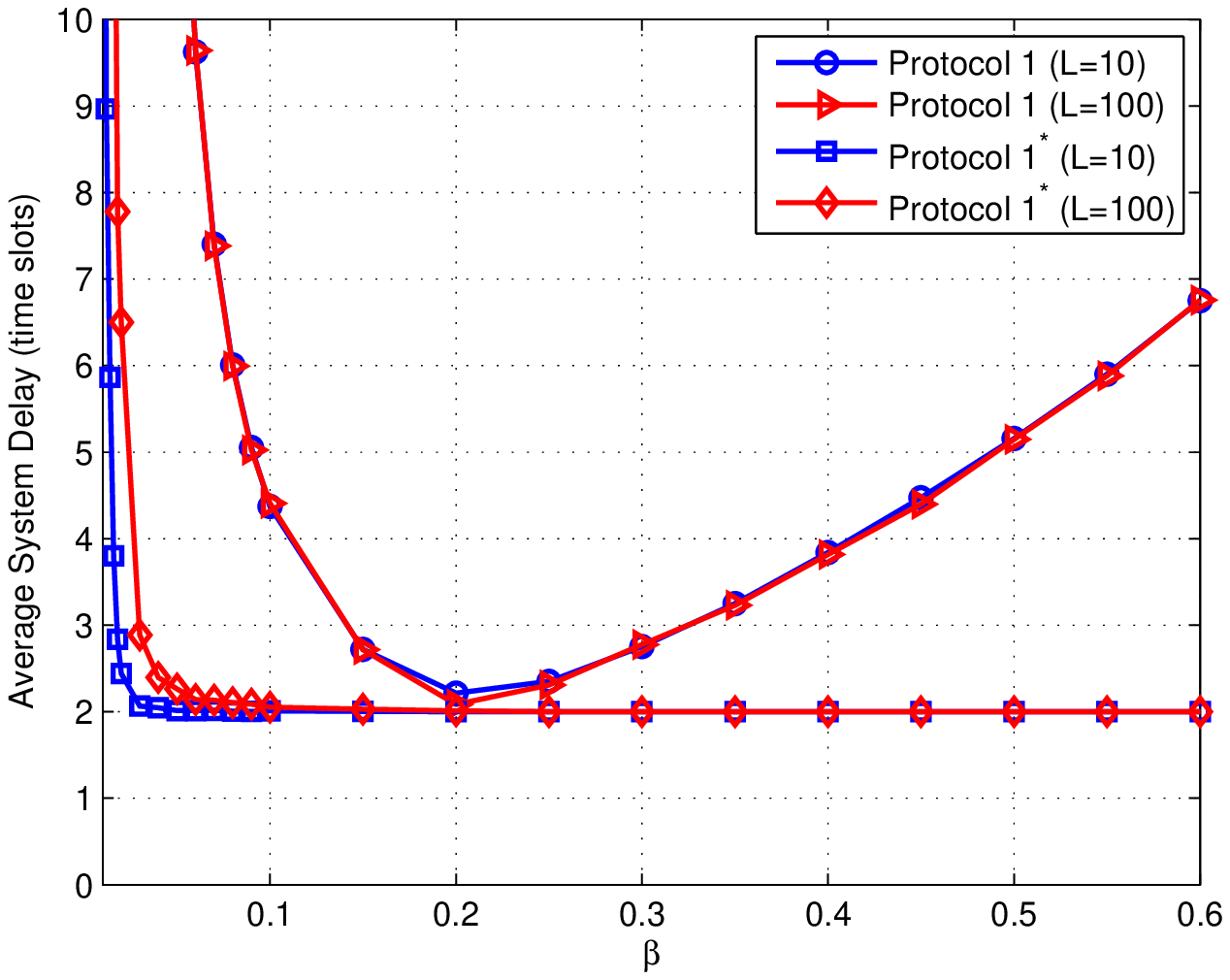}}
\subfigure[The effect of phase on the BER performance of Protocol~$1$]{ \label{fig:subfig:3e} 
\includegraphics[width=2.2in,height=1.8in]{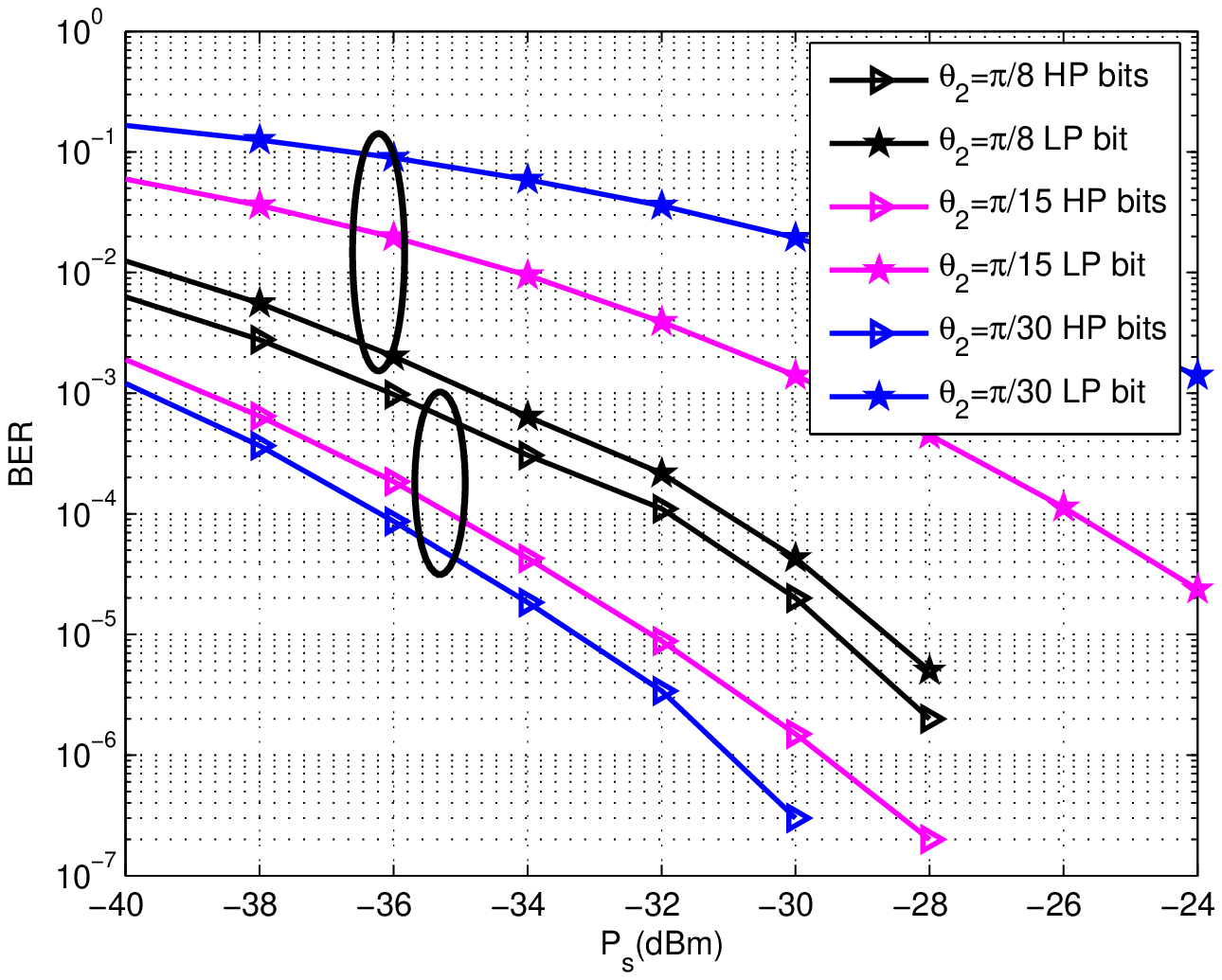}}
%
\caption{Some results for the buffer-aided single-relay WBAN.}
\label{fig:Fig.3} 
\end{figure*}

\section{Link Selection for Buffer-Aided Single-Relay WBAN}
In this section, we consider link-selection strategies for the buffer-aided single-relay case.

\subsection{Link-Selection Criterion}
In the conventional cooperative WBAN, no buffer is employed in the relay. In such a system, the basic transmission principle is described as follows. In the first time slot, the source node broadcasts the information to the relay and the destination. Then the relay performs error detection by using a cyclic-redundancy-check code. If the packet is not correctly detected, the relay doesn't transmit the decoded information to the destination; otherwise the relay transmits the information to the destination in the second time slot. Finally the destination obtains the estimated information depending on the $\textbf{S} \textbf{D}$ link or obtain the estimated information relying on both the $\textbf{S} \textbf{D}$ and $\textbf{R} \textbf{D}$ links (a maximum ratio combiner is used). It should be noted that the error detection at the relay is not considered in the buffer-aided relaying protocols.

Based on the link quality and the relay's buffer status information, the link selection criterion of a buffer-aided relaying WBAN system works as follows.
In the absence of a $\textbf{S} \textbf{D}$ link, suppose that the qualities of the $\textbf{S} \textbf{R}$  and $\textbf{R} \textbf{D}$ links in each time slot are $H_{\textbf{SR}}$ and $\beta H_{\textbf{RD}}$, respectively, where $\beta$ is a suitably chosen decision threshold and is a positive mnumber. Then, the Protocol~1 works as follows:
1)~If the buffer is neither empty nor full and the $\textbf{S} \textbf{R}$ link has the best quality, the source node transmits information to the relay node, otherwise the relay node transmits the information to the destination node;
2)~If the buffer is empty and the $\textbf{S} \textbf{R}$ link has the best link quality, the source node transmits information to the relay node, otherwise a silent time slot takes place;
3)~If the buffer is full and  the $\textbf{R} \textbf{D}$ link has the best quality, the relay node transmits the information to the destination node, otherwise a silent time slot takes place.

To facilitate practical WBAN applications, it is necessary to discuss the effects of the $\textbf{S} \textbf{D}$ link on BER performance in the proposed framework.
When considering the $\textbf{S} \textbf{D}$ link, the above protocol has to be modified. The basic idea is that the strongest link is selected for transmission or reception. The modified selection scheme (Modified Protocol~1) is described as:
1)~If the $\textbf{S} \textbf{D}$ link has the best quality, the source node transmits information to the destination node;
2)~If the buffer is not full and the $\textbf{S} \textbf{R}$ link has the best quality, the source node transmits information to the relay node;
3)~If the buffer is not empty and the $\textbf{R} \textbf{D}$ link has the best quality, the relay node transmits the information to the destination node;
4)~For other cases,  a silent time slot takes place.

For the Protocol~1, when the $\textbf{SR}$ ($\textbf{RD}$) link is repeatedly selected, a buffer at the relay can run full (empty). Moreover, if the quality of a link doesn't satisfy the constraint requirement, the link keeps idle, which leads to significant system delay. In other words, the average system delay of the Protocol~1 is caused by the queueing delay and the delay due to the silent time slots\footnote{The average system delay equals the ratio between the total number of time slots that the source node transmits all packets to the destination node and the total number of transmitted packets.}.
To address the above problem, by considering the relay's buffer status, Protocol~1 should be adjusted, which is referred to as Protocol~$1^{*}$. The detail of this protocol is illustrated as follows:
if the buffer is empty (full), one can select the $\textbf{SR}$ ($\textbf{RD}$) link without considering its quality; while the step 1) of the Protocol~1 is not changed. In contrast to Protocol~1, Protocol~$1^{*}$ has no silent time slot, i.e., its average system delay is only caused by the queueing delay. Hence, this protocol provides lower delay at the cost of sacrificing some BER performance since the best link is not always chosen.

Finally, to ensure reliable transmission of the HP bits for the medical information, an adaptive transmission scheme is proposed based on the Protocol~1.
To be specific, according to the link qualities, the source and relay nodes adopt dynamic HPSK modulations, where nonuniform phase parameters are set.
For example, an $8$-HPSK modulation with phase parameter $\bm{\theta}$ (i.e., $[\pi/4, \theta_2]$) is used in the source and relay nodes, where the first and second bits are HP bits while the third bit is LP bit.
For different link quality levels, a suitable element of $\theta_2$ is designed.
For the extremely poor link quality, we aim at ensuring reliable transmission for HP bits, while ignoring the reliability of the LP bit.
The detailed design and analysis of the proposed transmission scheme deserve a future work.


\begin{figure*}
\centering
\subfigure[BER Performance of different protocols]{ \label{fig:subfig:4a} 
\includegraphics[width=2.2in,height=1.8in]{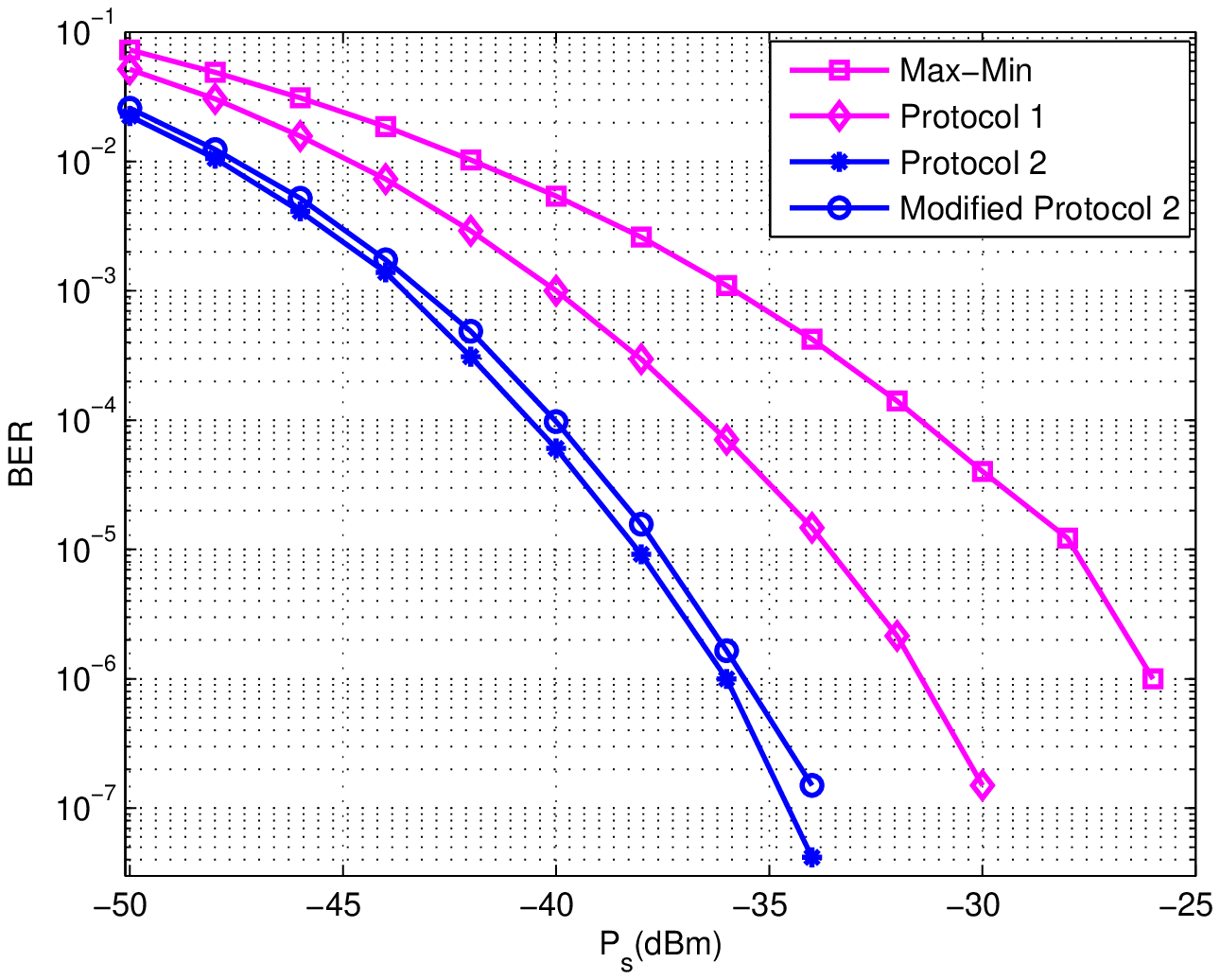}}
\subfigure[Average system delay of different protocols]{ \label{fig:subfig:4b} 
\includegraphics[width=2.2in,height=1.8in]{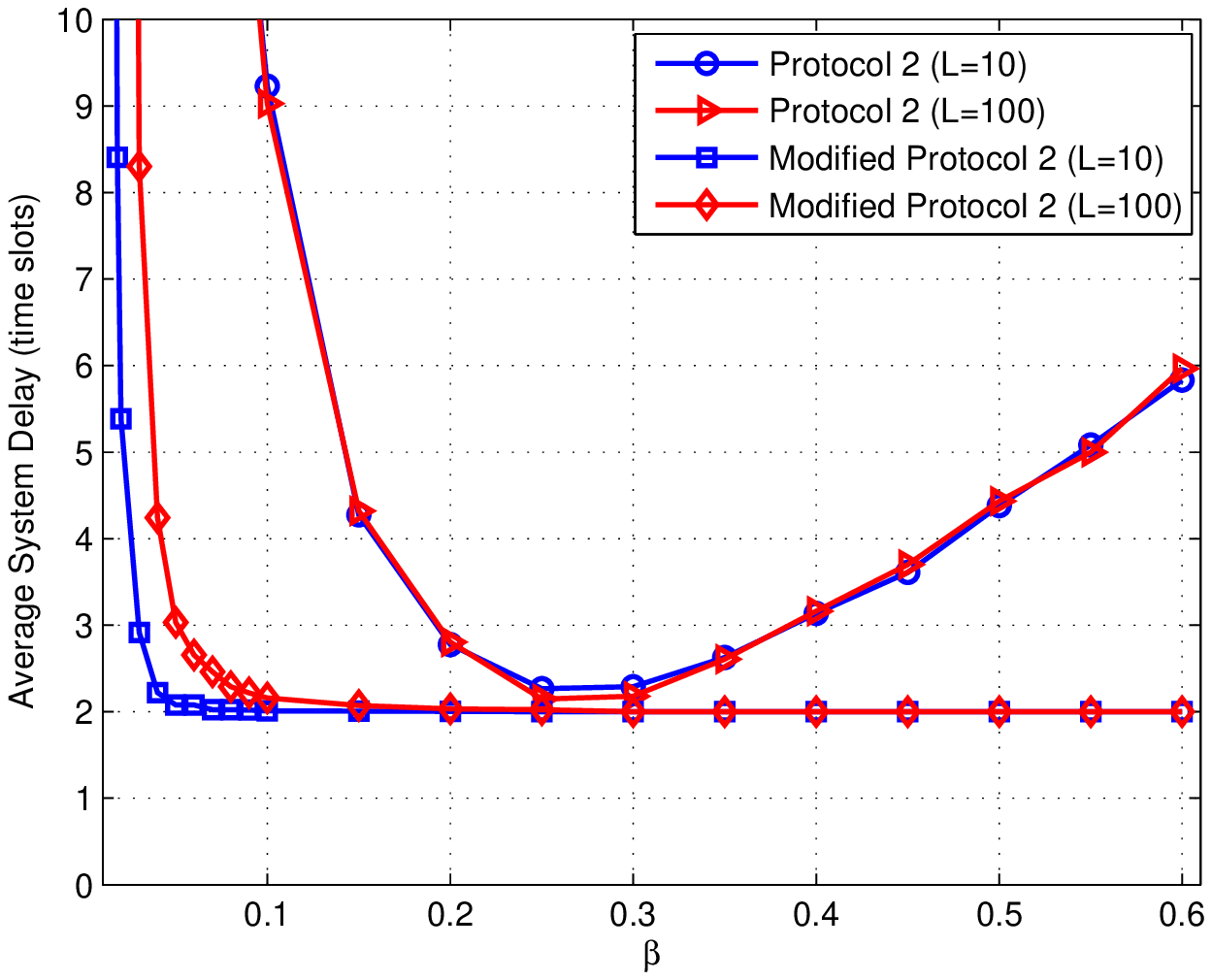}}
\subfigure[The effect of phase parameter on BER performance]{ \label{fig:subfig:4c} 
\includegraphics[width=2.2in,height=1.8in]{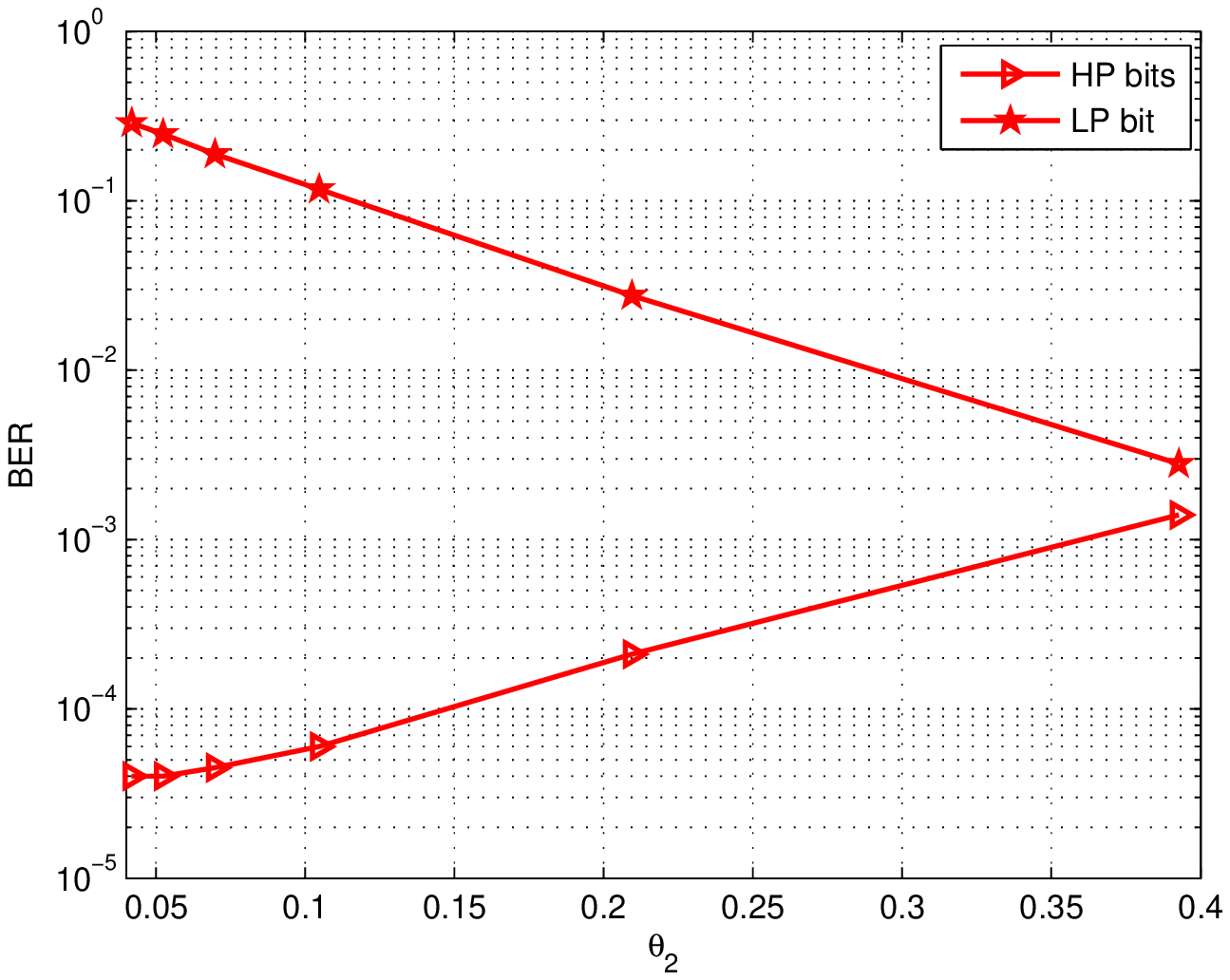}}
\vspace{-0.2cm}
\caption{Performance comparison of various protocols.}
\label{fig:Fig.4}  
\end{figure*}

\vspace{-2mm}
\subsection{Performance of the Proposed Link-Selection Strategy}
To explore the performance of the above protocols, we consider three transmission scenarios including six cases, as shown in Fig.~\ref{fig:subfig:3a}. In all simulations, we assume that the source node is embedded in the stomach at a depth $120$ mm, the maximum buffer size is $10$, the transmitted power for the relay is $-3$~dBm, the transmitted power for the source node is $P_s$, and QPSK modulation is adopted in all nodes.

Fig.~\ref{fig:subfig:3b} compares the BER performance among the six cases, where the decision threshold is set to $0.2$. We can observe that the proposed buffer-aided relaying WBAN can significantly save the energy consumption compared with the conventional cooperative and no cooperative WBANs. In particular, at a BER of $2 \times 10^{-6}$ the proposed system achieves $4-6$~dBm and $10$~dBm gains over the conventional cooperative and no cooperative WBANs, respectively.
Moreover, the Modified Protocol~1 can provide better performance than that of the Protocol~1. However, with increasing the distance of the $\textbf{SD}$ link, the performance gain disappears. For example, compared with the case~2/5, at a BER of $10^{-7}$, the case~3 has a 2~dBm gain while the case~4 only provides a 0.2~dBm gain.
Therefore, the target hub is generally embedded into the mobile phone or PDA and can be flexibly moved, thus the $\textbf{SD}$ link is not considered in practical WBAN applications (i,e., the case~5 will be considered in the following simulation.).

Fig.~\ref{fig:subfig:3d} and \ref{fig:subfig:3d-1} show BER performance of the Protocol~1 and Protocol~$1^{*}$ with different values of decision threshold and maximum buffer size, respectively. Referring to these two figures, it is obvious that,
1)~the Protocol~1 achieves better BER performance with the increasing of decision threshold, while it cannot do so with the increasing of maximum buffer size;
2)~BER performance of the Protocol~$1^{*}$ depends on both the values of decision threshold and maximum buffer size.
Moreover, Fig.~\ref{fig:subfig:3d-2} presents the average system delay for different values of decision threshold for the Protocol~1 and Protocol~$1^{*}$. It can be observed that for the Protocol~1 there exists an optimal value of decision threshold (i.e., 0.2) to minimize the average system delay.
In addition, the Protocol~$1^{*}$ provides lower average system delay than the Protocol~1.
For example, when decision threshold is $0.2$ and maximum buffer size is $10$, the Protocol~1 requires a delay of $2.2$ time slots while the Protocol~$1^{*}$ only needs $2$ time slots.
Hence, the decision threshold should be adjusted to strike a good balance between the average delay and BER performance for the Protocol~1, while the decision threshold and buffer size should be jointly adjusted to satisfy the BER requirements for the Protocol~$1^{*}$.

To satisfy various QoS requirements for different types of medical information, the phase parameter should be adjusted. Fig.~\ref{fig:subfig:3e} presents the performance of the Protocol~1 with $8$-HPSK modulation for different values of $\theta_2$, i.e., $\pi/8$, $\pi/15$, $\pi/30$, where the first and second bits are HP bits and the third bit is the LP bit. According to this figure, the system is able to achieve different BER performance (i.e., satisfying different QoS requirement) by adjusting the phase parameter. For example, at a BER of $2 \times 10^{-6}$, the HP bits with $\pi/30$ has a $3.5$~dBm gain over that with $\pi/8$. This gain can be used to resist the deeply fading.

\section{Relay Selection for Buffer-Aided Multi-Relay WBAN}
Here, we consider relay-selection strategies for the buffer-aided multiple-relay WBAN, in which the $\textbf{SD}$ link is not available.

\vspace{-3mm}
\subsection{Relay Selection Criterion}
In the conventional cooperative DF-based multi-relay WBAN without buffers, the {\it{max-min}} relay-selection strategy is considered as the optimal selection scheme, which is viewed as the baseline relay-selection strategy. 


Based on the Protocol~1, a new relay selection strategy (Protocol~2) for the proposed framework is developed, which can be also referred as to max-all-link selection protocol. The basic idea of the Protocol~2 is that the link having the best quality is selected to transmit or receive information according to the status of the relays' buffers. More specifically, for a given $\textbf{SR}$ link, the relay can receive information from the source node if its buffer is not full, while for a given $\textbf{RD}$ link, the relay can forward the source-node's information towards the destination node if the relay node is not empty. Similar to the Protocol~1, the Protocol~2 also suffers from a significant system delay. For example, when all relays are sitting in empty state, the $\textbf{RD}$ links always have the best quality, thus resulting in idle state of the system.

To overcome the above weakness, similarly to the Protocol~1*, we improve the Protocol~2 by taking the status of the relays' buffers into consideration so as to formulate the Modified Protocol~2, which can be also referred as to max-link selection protocol. The principle of the Modified Protocol~2 is that based on the the status of the relays' buffers the available links are compared to find the most reliable link, which is then used to transmit information\footnote{It should be noted that the link from the source node to the relay node is considered as an available link when the buffer status of the corresponding relay node is not full, while the link from the relay node to the destination node is considered as an available link when the buffer status of the corresponding relay node is not empty.}.
Through the above operation, the system is always working in transmitted or received status for every time slot and is not sitting idle status, thus significantly reducing the system delay.

\vspace{-2mm}
\subsection{Performance of the Proposed Relay-Selection Strategy}
Fig.~\ref{fig:subfig:4a} depicts the BER performance of various protocols (\textit{max-min}, Protocol~1, Protocol~2, and Modified Protocol~2) vs. transmit power $P_s$ of the source node. We consider a buffer-aided two-relay WBAN, and assume that the distances between the source node and each relay node, and between each relay node and target hub are $120$ mm, the buffer size is $10$, the transmitted power for each relay node is $-6$~dBm, the decision threshold is 0.2, and QPSK modulation is adopted. As illustrated in this figure, the Protocol~2 and the Modified Protocol~2 achieve a significant gain with respect to the max-min protocol and Protocol~1. In addition, the Modified Protocol~2 has a $0.4$~dBm performance loss compared with the Protocol~2 at a BER of $10^{-6}$.
Moreover, in Fig.~\ref{fig:subfig:4b}, we show the average system delay of the Protocol~2 and Modified Protocol~2 with different buffer sizes vs. threshold decision $\beta$. It can be seen that although the Modified Protocol~2 provides slightly worse BER performance than the Protocol~2, it significantly reduces the average system delay because no silent time slot is involved.
For example, when the decision threshold is about $0.25$ and the maximum buffer size is $10$, the Protocol $2$ requires a delay of $2.26$ time slots while the Modified Protocol $2$ only needs $2$ time slots.
Furthermore, the proposed adaptive hierarchical-modulation scheme can be easily extended to the Protocol~2 and Modified Protocol~2.
Fig.~\ref{fig:subfig:4c} shows the effect of phase parameter on BER performance, where the Protocol~2 is employed and $P_s$ is set to $-40$~dBm. As seen, with the decrease of the value of $\theta_2$ the HP bits can be transmitted reliably at the cost of decreasing the performance of the LP bit. Hence, in the extremely poor channel link, we can guarantee the reliability for the HP bits rather than the LP bits by adjusting the phase parameter.

\section{Challenges and Future Research Directions}\label{sect:FDD}
Through the above discussion, the proposed buffer-aided multi-relaying implant WBAN framework can provide excellent BER performance, and hence it can be considered as a promising candidate for healthcare IoT.  Here, several interesting problems and practical challenges on this research area are discussed.

{\bf{\textit{1)~BER, Throughput and Delay Analysis}}}:
Unlike the through-the-air wireless channel, the lognormal-distribution-based channel is considered in the WBAN. It is intuitive that under the buffer-aided multi-relaying WBAN the performance analysis become more difficult. Consequently, BER, throughput and delay analysis for the proposed framework is a challenging issue especially considering the HPSK modulation.

{\bf{\textit{2)~Adaptive Hierarchical Modulation}}}:
In the proposed framework, QoS problem is achieved through the usage of hierarchical modulations, where the HP and LP medical information are simultaneously transmitted to mitigate the system delay.
To ensure reliable transmission of the HP medical information and minimal impact on LP medical information transmission, adaptive hierarchical modulation transmission strategies deserve further investigation.

{\bf{\textit{3)~Low-Complexity and Low-Delay Relay Selection}}}: Under the proposed framework, the Modified Protocol~2 has been proposed to significantly reduce the system delay at the expense of a degradation of BER performance.
However, the implementation complexity and system delay of this protocol increase as the number of relays becomes larger. Hence, developing simplified relay-selection strategies is expected.

{\bf{\textit{4)~Energy Efficiency Investigation}}}:  The design goal of the implant WBAN is that minimum energy consumption is used to reliably transmit medical information. In contrast to the conventional cooperative WBAN, the system delay is a critical factor for the proposed framework. Consequently, how to build an energy-efficient analytical model for the proposed framework and investigate the tradeoff among energy consumption, BER and system delay is an important issue.

{\bf{\textit{5)~Physical-Layer Security}}}: The buffer-aided multi-relaying networks can greatly boost the
security performance. In the proposed framework, the implant WBAN simultaneously supports delay-sensitive and delay-tolerant information. Nonetheless, for practical delay-sensitive and delay-tolerant applications, how to achieve a balance between the system security and the system delay is required to be investigated in the future.

\section{Conclusions}\label{sect:conc}
In this article, an overview of conventional cooperative WBAN framework in healthcare IoT has been provided. To address the drawbacks of the conventional cooperative WBAN, a QoS-aware buffer-aided relaying framework has been proposed for implant WBAN. Moreover, some transmission strategies of the buffer-aided single- and multi-relay implant WBAN systems have been developed. As a further insight, the BER performance and system delay of the proposed implant WBAN have been investigated. Simulated results have illustrated that the proposed framework offers a significant performance gain compared with the conventional cooperative WBAN. Hence, the proposed framework can be viewed as an efficient approach for the implant WBAN. As an emerging research area for healthcare IoT, there are a variety of open problems related to the buffer-aided relaying WBAN awaiting further investigation, some of which have been listed in Sect.~V.

\section*{Acknowledgements}
This work was supported in part by the NSF of China (Nos. 61701121, 61771149, 11871248, 61871136),
the Open Research Fund of State Key Laboratory of Integrated Services Networks under Grant ISN19-04, the Guangdong Province Universities and Colleges Pearl River Scholar Funded Scheme under Grant 2017-ZJ022, the Research Projects of the Education Department of Guangdong Province (Nos. 2017KTSCX060, 2017KZDXM028), and the Fundamental Research Funds for the Central Universities under Grant 21618329.

\section*{Biographies}

\noindent \footnotesize{Guofa Cai (caiguofa2006@gmail.com) received the Ph.D. degree in communication engineering from Xiamen University, Xiamen, China, in 2015. In 2017, he was a Research Fellow at the School of Electrical and Electronic Engineering, Nanyang Technological University, Singapore. He is currently an Associate Professor with the School of Information Engineering, Guangdong University of Technology, China. His primary research interests include information theory and coding, spread-spectrum modulation, wireless body area networks, and Internet of Things.}

\vspace{0.5cm}

\noindent Yi Fang (fangyi@gdut.edu.cn) received the Ph.D. degree in communication engineering, Xiamen University, China, in 2013. From February 2014 to February 2015, he was a Research Fellow at the School of Electrical and Electronic Engineering, Nanyang Technological University, Singapore. He is currently an Associate Professor at the School of Information Engineering, Guangdong University of Technology, China. His research interests include information and coding theory, spread-spectrum modulation, and cooperative communications. He is the corresponding author of this paper.

\vspace{0.5cm}

\noindent Jinming Wen (jinming.wen@mail.mcgill.ca) is a full professor in the College of Information Science and Technology and the College of Cyber Security, Jinan University, Guangzhou. His research interests are in the areas of lattice reduction and sparse recovery. He has published around 45 papers in top journals (including Applied and Computational Harmonic Analysis, IEEE Transactions on Information Theory/ Signal Processing/Wireless Communications) and conferences.
He is an Associate Editor of IEEE Access.

\vspace{0.5cm}

\noindent Guojun Han (gjhan@gdut.edu.cn) obtained his Ph.D. from Sun Yat-sen University, Guangzhou, China. From March 2011 to August 2013, he was a Research Fellow at the School of Electrical and Electronic Engineering, Nanyang Technological University, Singapore. From October 2013 to April 2014, he was a Research Associate at the Department of Electrical and Electronic Engineering, Hong Kong University of Science and Technology. He is now a Full Professor and Dean at the School of Information Engineering, Guangdong University of Technology, Guangzhou, China. His research interests include wireless communications, coding and signal processing for data storage.
\vspace{0.5cm}

\noindent Xiaodong Yang (xdyang@xidian.edu.cn) is an Associate Professor with the School of Electronic Engineering, XiDian University, Xi'an, China. He has published over 70 papers in peer-reviewed journals. His main research area is Body Area Networks. He received the Young Scientist Award from the International Union of Radio Science in 2014. He is on the editorial board of several IEEE and IET journals, including IEEE Journal of Electromagnetics, RF and Microwave in Medicine and Biology, etc. He has a global collaborative
research network in the related fields. He is a Senior Member of IEEE.

\end{document}